\RequirePackage{fix-cm}
\documentclass[smallextended]{svjour3}       % onecolumn (second format)
\smartqed  % flush right qed marks, e.g. at end of proof
\usepackage{graphicx}
\usepackage{multirow}
\usepackage{epsfig}
\journalname{}

\title{Parity proofs of the Kochen-Specker theorem based on 60 complex rays in four dimensions}

\author{Mordecai Waegell and P.K. Aravind }

\authorrunning{M.Waegell, P.K. Aravind}

\institute{M.Waegell, P.K. Aravind \at
Physics Department, Worcester Polytechnic Institute, Worcester, MA 01609, U.S.A.\\
\email{caiw@wpi.edu, paravind@wpi.edu}}

\date{\today}

\begin{document}
\maketitle
\begin{abstract}
 It is pointed out that the 60 complex rays in four dimensions associated with a system of two qubits yield over $10^{9}$ critical parity proofs of the Kochen-Specker theorem. The geometrical properties of the rays are described, an overview of the parity proofs contained in them is given, and examples of some of the proofs are exhibited.\\\\
\end{abstract}

\section{\label{sec:Intro}Introduction}

In a recent paper \cite{Waegell2010} we showed that the 60 real rays in four dimensions derived from the vertices of a 600-cell yield over hundred million critical parity proofs of the Kochen-Specker (KS) theorem \cite{KS1967}. In the present paper we point out that the 60 complex rays in four dimensions connected with a pair of qubits lead to a similarly large number of parity proofs. Both these 60-ray systems can be regarded as generalizations, in different ways, of the set of 24 real rays in four dimensions used by Peres \cite{Peres1991} and others \cite{Kernaghan1994,Cabello1996,Aravind2000,Pavicic2010,Waegell2011a} to give proofs of the KS theorem.\\

The real and complex systems of 60 rays have 12 rays in common, but do not share a single parity proof. The proofs in both systems are interesting for several reasons: they are very transparent and take no more than simple counting to verify; they are extremely numerous, numbering over a 100 million in each case; and they possess many geometrical features of interest \cite{BengtssonBook}. In addition to these purely mathematical reasons, they are of physical interest because they can be converted into experimentally testable Bell inequalities, as pointed out in \cite{Aolita,Cabello2010}, and also tie in with discussions of contextuality \cite{Klyachko,Liang,Bengtsson} and nonlocality \cite{Cabello2008} in four-state systems. Four-state systems provide an attractive setting for a discussion of these ideas because they can be realized experimentally using ions \cite{Kirchmair}, neutrons \cite{Bartosik}, photons \cite{Amselem} and nuclear spins \cite{Moussa}. The present parity proofs could also find application in protocols such as quantum key distribution \cite{BPPeres}, random number generation \cite{Svozil} and parity oblivious transfer \cite{Spekkens}.\\

The plan of this paper is as follows. Sec.~\ref{sec:60-105} introduces the 60 rays and 105 bases connected with a system of two-qubits and points out their important properties. It also explains what a ``parity proof" is (as well as the notion of a {\it critical} parity proof) and introduces the two alternative symbols we use for such proofs. Sec.~\ref{sec:40-40} introduces a 40-40 subsystem (i.e., one consisting of 40 rays and 40 bases) of the 60-105 system and explores its parity proofs. The proofs are of just six different sizes (although each consists of a number of geometrically distinct varieties), but their replicas under symmetry run into the thousands. We point out that a projective configuration originating in the work of Desargues holds the key to a geometrical construction of all the proofs of the smallest size. Sec.~\ref{sec:36-36} introduces a 36-36 subsystem whose parity proofs are more varied than those of the 40-40 system, but which can also be completely delineated. Sec.~\ref{sec:Full} discusses a variety of other interesting parity proofs in the 60-105 system. Finally, Sec.~\ref{sec:Discussion} contains some concluding remarks.

\section{\label{sec:60-105}The 60 complex rays and their bases}

A system of two qubits has 15 nontrivial observables, which can be taken to be the Pauli operators of the qubits and their direct products. These 15 observables form 15 triads of commuting observables that are shown in the first column of Table \ref{60States}. The simultaneous eigenstates of these triads give rise to the 60 rays (as we will term the eigenstates) shown in the second column of Table \ref{60States}. These 60 rays form the 105 bases shown in Table \ref{tab:Bases}, with the top 15 already being present in Table \ref{60States}. This 60-105 system  provides the grand framework within which all the parity proofs of this paper are embedded. Table \ref{tab:Bases} has the following important properties:\\ \\(1) It contains two different types of bases: 15 ``pure" bases that arise as eigenstates of the triads in Table \ref{60States} (and shown in the top part of the table), and 90 ``hybrid" bases that arise through a mating (or hybridization) of a pair of pure bases. For example, the hybrid bases 1 2 15 16 and 3 4 13 14 arise through a mating of the pure bases 1 2 3 4 and 13 14 15 16 and consist of equal (and complementary) mixes of rays from the two of them. \\ \\(2) Each ray occurs in exactly seven bases. In addition to the brief symbol 60-105 we have introduced for this system, we will sometimes use the expanded symbol $60_{7}$-$105_{4}$ to indicate that each of the rays occurs in 7 bases (the subscript 4 simply indicates that each of the bases consists of four rays, a fact which is true of all the configurations to be discussed in this paper). Note the check $60\times7=105\times4$ on the numbers entering the expanded symbol.\\ \\(3) Each ray is orthogonal to 15 others and occurs three times with three of them and once each with the twelve others in the seven bases it occurs in. The fact that all the orthogonalities between the rays are represented in their basis table is sometimes conveyed by saying that the system is {\bf saturated}. For a saturated system, such as the present one, the basis table is completely equivalent to the Kochen-Specker diagram of the rays. \\ \\(4) The 60-105 system contains (in ten different ways) the 24-24 system of rays and bases used by Peres and others to give proofs of the KS theorem. However it also contains an (astronomically!) large number of new proofs that cannot be reduced to those of the Peres system simply by shedding rays or bases. It is these new proofs that are the subject of this paper.\\ \\(5) The symmetry group of the 60-105 system (i.e. the group of unitary transformations that keeps its rays and bases invariant as a whole) has 11,520 elements in it. This can be seen as follows. Each symmetry operation maps a particular pure basis, say 1 2 3 4, into any of the pure bases, including itself, and so is described by an operator of the form $U = \mid x><1\mid + a\mid y><2\mid + b\mid z><3\mid + c\mid w><4\mid$, where $x$ $y$ $z$ $w$ is the basis into which the mapping occurs. However all 24 permutations of the final basis states are allowed and the phase factors $a,b$ and $c$ can each take on the values $\pm 1,\pm i$ subject to the restriction that only an even number of $i's$ occur. The order of the symmetry group is therefore $15\times24\times32 = 11,520.$\\

\begin{table}[ht]
\centering % used for centering table
\begin{tabular}{|c | c c c c |} % centered columns (4 columns)
\hline % inserts single horizontal line
Operators & ++  &   + --  &  -- +  &  -- --  \\
\hline
$Z_1$, $Z_2$, $Z_1$$Z_2$            & $1=1000$ & $2=0100$ & $3=0010$ & $4=0001$ \\
$X_1$, $X_2$, $X_1$$X_2$            & $5=1111$ & $6=1\overline{1}1\overline{1}$ & $7=11\overline{1}\overline{1}$ & $8=1\overline{1}\overline{1}1$ \\
$Y_1$, $Y_2$, $Y_1$$Y_2$            & $9=1ii\overline{1}$ & $10=1\overline{i}i1$ & $11=1i\overline{i}1$ & $12=1\overline{i}\overline{i}\overline{1}$ \\
$Z_1$, $X_2$, $Z_1$$X_2$            & $13=1100$ & $14=1\overline{1}00$ & $15=0011$ & $16=001\overline{1}$ \\
$X_1$, $Y_2$, $X_1$$Y_2$            & $17=1i1i$ & $18=1\overline{i}1\overline{i}$ & $19=1i\overline{1}\overline{i}$ & $20=1\overline{i}\overline{1}i$ \\
$Y_1$, $Z_2$, $Y_1$$Z_2$            & $21=10i0$ & $22=010i$ & $23=10\overline{i}0$ & $24=010\overline{i}$ \\
$Z_1$, $Y_2$, $Z_1$$Y_2$            & $25=1i00$ & $26=1\overline{i}00$ & $27=001i$ & $28=001\overline{i}$ \\
$X_1$, $Z_2$, $X_1$$Z_2$            & $29=1010$ & $30=0101$ & $31=10\overline{1}0$ & $32=010\overline{1}$ \\
$Y_1$, $X_2$, $Y_1$$X_2$            & $33=11ii$ & $34=1\overline{1}i\overline{i}$ & $35=11\overline{i}\overline{i}$ & $36=1\overline{1}\overline{i}i$ \\
$Z_1$$X_2$, $X_1$$Z_2$, $Y_1$$Y_2$  & $37=111\overline{1}$ & $38=11\overline{1}1$ & $39=1\overline{1}11$ & $40=1\overline{1}\overline{1}\overline{1}$ \\
$X_1$$Y_2$, $Y_1$$X_2$, $Z_1$$Z_2$  & $41=100i$ & $42=01\overline{i}0$ & $43=01i0$ & $44=100\overline{i}$ \\
$Y_1$$Z_2$, $Z_1$$Y_2$, $X_1$$X_2$  & $45=1ii1$ & $46=1\overline{i}i\overline{1}$ & $47=1i\overline{i}\overline{1}$ & $48=1\overline{i}\overline{i}1$ \\
$Z_1$$Z_2$, $X_1$$X_2$, $-Y_1$$Y_2$ & $49=1001$ & $50=100\overline{1}$ & $51=0110$ & $52=01\overline{1}0$ \\
$Z_1$$X_2$, $X_1$$Y_2$, $-Y_1$$Z_2$ & $53=11\overline{i}i$ & $54=11i\overline{i}$ & $55=1\overline{1}ii$ & $56=1\overline{1}\overline{i}\overline{i}$ \\
$Z_1$$Y_2$, $X_1$$Z_2$, $-Y_1$$X_2$ & $57=1i1\overline{i}$ & $58=1i\overline{1}i$ & $59=1\overline{i}1i$ & $60=1\overline{i}\overline{1}\overline{i}$ \\
\hline
\end{tabular}
\caption{The 15 triads of commuting observables for a pair of qubits and their simultaneous eigenstates. $X_{i},Y_{i},Z_{i}, (i=1,2)$ are the Pauli operators of the two qubits. The eigenstates, or rays, are numbered from 1 to 60. The components of the (unnormalized) rays are given in an orthonormal basis, with commas omitted between components and a bar over a number indicating its negative. The eigenvalue signatures of the eigenstates with respect to the first two observables in the defining triad are shown at the top (the signature for the third observable is always such as to make the product of the three signatures a +)}
\label{60States} % is used to refer this table in the text
\end{table}

\begin{table}[ht]
\centering % used for centering table
\begin{tabular}{|c c c c | c c c c | c c c c | c c c c | c c c c|} % centered columns (4 columns)
\hline % inserts single horizontal line
1 & 2 & 3 & 4 & 5 & 6 & 7 & 8 & 9 & 10 & 11 & 12 & 13 & 14 & 15 & 16 & 17 & 18 & 19 & 20 \\
21 & 22 & 23 & 24 & 25 & 26 & 27 & 28 & 29 & 30 & 31 & 32 & 33 & 34 & 35 & 36 & 37 & 38 & 39 & 40 \\
41 & 42 & 43 & 44 & 45 & 46 & 47 & 48 & 49 & 50 & 51 & 52 & 53 & 54 & 55 & 56 & 57 & 58 & 59 & 60 \\
\hline
1 & 2 & 15 & 16 & 1 & 2 & 27 & 28 & 1 & 3 & 22 & 24 & 1 & 3 & 30 & 32 & 1 & 4 & 42 & 43 \\
1 & 4 & 51 & 52 & 2 & 3 & 41 & 44 & 2 & 3 & 49 & 50 & 2 & 4 & 21 & 23 & 2 & 4 & 29 & 31 \\
3 & 4 & 13 & 14 & 3 & 4 & 25 & 26 & 5 & 6 & 19 & 20 & 5 & 6 & 31 & 32 & 5 & 7 & 14 & 16 \\
5 & 7 & 34 & 36 & 5 & 8 & 46 & 47 & 5 & 8 & 50 & 52 & 6 & 7 & 45 & 48 & 6 & 7 & 49 & 51 \\
6 & 8 & 13 & 15 & 6 & 8 & 33 & 35 & 7 & 8 & 17 & 18 & 7 & 8 & 29 & 30 & 9 & 10 & 23 & 24 \\
9 & 10 & 35 & 36 & 9 & 11 & 18 & 20 & 9 & 11 & 26 & 28 & 9 & 12 & 38 & 39 & 9 & 12 & 49 & 52 \\
10 & 11 & 37 & 40 & 10 & 11 & 50 & 51 & 10 & 12 & 17 & 19 & 10 & 12 & 25 & 27 & 11 & 12 & 21 & 22 \\
11 & 12 & 33 & 34 & 13 & 14 & 27 & 28 & 13 & 15 & 34 & 36 & 13 & 16 & 39 & 40 & 13 & 16 & 55 & 56 \\
14 & 15 & 37 & 38 & 14 & 15 & 53 & 54 & 14 & 16 & 33 & 35 & 15 & 16 & 25 & 26 & 17 & 18 & 31 & 32 \\
17 & 19 & 26 & 28 & 17 & 20 & 43 & 44 & 17 & 20 & 54 & 56 & 18 & 19 & 41 & 42 & 18 & 19 & 53 & 55 \\
18 & 20 & 25 & 27 & 19 & 20 & 29 & 30 & 21 & 22 & 35 & 36 & 21 & 23 & 30 & 32 & 21 & 24 & 47 & 48 \\
21 & 24 & 53 & 56 & 22 & 23 & 45 & 46 & 22 & 23 & 54 & 55 & 22 & 24 & 29 & 31 & 23 & 24 & 33 & 34 \\
25 & 28 & 46 & 48 & 25 & 28 & 59 & 60 & 26 & 27 & 45 & 47 & 26 & 27 & 57 & 58 & 29 & 32 & 38 & 40 \\
29 & 32 & 58 & 60 & 30 & 31 & 37 & 39 & 30 & 31 & 57 & 59 & 33 & 36 & 42 & 44 & 33 & 36 & 57 & 60 \\
34 & 35 & 41 & 43 & 34 & 35 & 58 & 59 & 37 & 38 & 55 & 56 & 37 & 39 & 58 & 60 & 37 & 40 & 49 & 52 \\
38 & 39 & 50 & 51 & 38 & 40 & 57 & 59 & 39 & 40 & 53 & 54 & 41 & 42 & 54 & 56 & 41 & 43 & 57 & 60 \\
41 & 44 & 51 & 52 & 42 & 43 & 49 & 50 & 42 & 44 & 58 & 59 & 43 & 44 & 53 & 55 & 45 & 46 & 53 & 56 \\
45 & 47 & 59 & 60 & 45 & 48 & 50 & 52 & 46 & 47 & 49 & 51 & 46 & 48 & 57 & 58 & 47 & 48 & 54 & 55 \\
\hline
\end{tabular}
\caption{Basis table of the 60 rays of Table \ref{60States}. The rays form 105 bases, with each ray occurring in 7 bases. The 15 pure bases are shown in the top section, and the 90 hybrid bases below.}
\label{tab:Bases} % is used to refer this table in the text
\end{table}

We now explain what we mean by a parity proof. A parity proof of the KS theorem in $d$ dimensions, with $d$ even, is a set of $R$ rays and $B$ bases, with $B$ odd, with the property that each of the $R$ rays occurs an even number of times among the $B$ bases. Such a set provides a proof of the KS theorem because it is impossible to assign 0/1 values to the rays, in a noncontextual fashion, in such a way that each basis has one 1 and $d-1$ 0's in it. The even multiplicity of the rays, together with the odd number of bases, makes such an assignment impossible and it is this even-odd contradiction that gives the parity proof its name. The parity proofs presented in this paper will all be in dimension $d = 4$.\\

A parity proof involving $R$ rays and $B$ bases will be denoted by the symbol $R$-$B$. However this symbol is not very informative because it does not reveal the multiplicities (i.e. the number of occurrences) of the rays in the proof. This defect can be remedied by using the expanded symbol $R^{'}_{m'}R^{''}_{m''}\ldots-B_{d}$, which reveals that there are $R^{'}$ rays of multiplicity $m^{'}$, $R^{''}$ rays of multiplicity $m^{''}$, etc., in the $B$ bases of the proof (each of which contains $d$ rays). The foregoing numbers are not all independent but obey the constraint $m^{'}R^{'} + m^{''}R^{''} + \ldots = d\cdot B$. The expanded symbol makes it easy to check the validity of a parity proof. Consider a parity proof involving 40 rays and 25 bases with the expanded symbol $32_{2}6_{4}2_{6}$-$25_{4}$. This symbol reveals that there are 32 rays of multiplicity two, 6 rays of multiplicity four and 2 rays of multiplicity six, which together account for the 100 rays in the 25 bases of the proof. We will usually use the brief symbol for a parity proof and resort to the expanded one only when the additional detail proves helpful. We should mention that we will sometimes use the brief and expanded symbols not for parity proofs but for systems of rays and bases housing parity proofs. For example, we have already used the symbols 60-105 and $60_{7}$-$105_{4}$ to refer to the system of rays and bases within which all our parity proofs are embedded. It should be obvious from the context whether a symbol refers to a parity proof or to a system of rays and bases housing parity proofs.\\

The only parity proofs we will present in this paper are ones that are {\bf basis-critical}. A $R$-$B$ proof will be said to be basis-critical if dropping even a single basis from it makes it possible to assign noncontextual 0/1 values to the rays in the surviving $B-1$ bases in such a way that each basis has one 1 and $d-1$ 0's in it. Basis-critical proofs are of interest because they permit the tightest experimental tests of the KS theorem. Proofs that are not basis-critical can always be reduced to ones that are by shedding bases, and so we will ignore them. The use of basis-criticality as a filter allows us to pare down our list of proofs considerably by eliminating redundant proofs that contain smaller proofs within them.

\section{\label{sec:40-40}The 40-40 subsystem and its parity proofs}

The 60-105 system has six 40-40 subsystems in it. These subsystems are of interest because they have a simple pattern of parity proofs. We explain how to obtain these subsystems, and then proceed to extract all the parity proofs in one of them.\\

The 15 triads of observables in Table \ref{60States} can be grouped into six sets of five (see Table \ref{tab:MUBs}) in such a way that each set defines a system of mutually unbiased bases (MUBs) \cite{Durt}. Two bases are said to be mutually unbiased if the absolute value of the inner product of any normalized ray of one with any normalized ray of the other is always the same (and equal to $\frac{1}{2}$ in dimension 4). The maximum number of MUBs in dimension 4 is five. The triads in each row of Table \ref{tab:MUBs} define a maximal set of MUBs, with any two rows having one triad (and therefore one basis) in common. A $40_{4}$-$40_{4}$ system can be obtained by dropping the 20 rays defined by the triads in any row of Table \ref{tab:MUBs} from the full set of 60 rays and keeping only the 40 bases made up exclusively of the remaining 40 rays. This construction can be described by the schematic equation\\

$(60_{7}$-$105_{4})$ - (a maximal set of MUBs) =  $40_{4}$-$40_{4}$       (in 6 different ways) \\

\begin{figure}[htp]
\begin{center}
\includegraphics[width=1.00\textwidth]{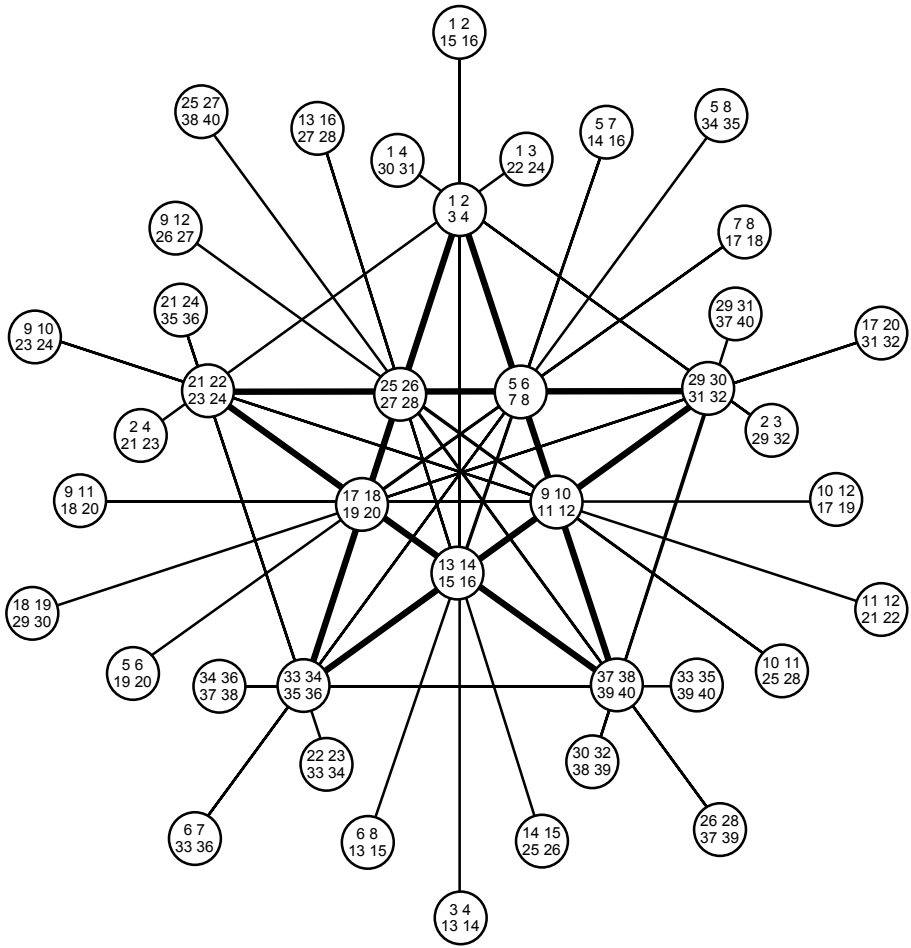}
\end{center}
\caption{The 40 bases formed by the 40 rays of Table \ref{40States}. Each ray occurs in exactly four bases, so this is a $40_{4}$-$40_{4}$ system. The ten bases closest to the center are pure bases, while the remaining thirty are hybrids. The four pure bases along any edge of the central pentagram are mutually unbiased. The pure bases mate in pairs to produce all the hybrids. The lighter lines pass through a pair of pure bases and the two hybrids they give rise to.}
\label{tab:40Bases}
\end{figure}

Table \ref{40States} shows the 40 rays that result if one drops the triads in the last row of Table \ref{tab:MUBs}, and Fig. \ref{tab:40Bases} shows the 40 bases formed by these rays. {\it Note that the numbering of the rays in Table \ref{40States} and Fig. \ref{tab:40Bases} is different from that in the earlier sections, and it is this new numbering that will be used throughout this section.}\\

\begin{table}[ht]
\centering % used for centering table
\begin{tabular}{|c | c c c c c |} % centered columns (4 columns)
\hline % inserts single horizontal line
1 & \{$Z_1$, $Z_2$, $Z_1$$Z_2$\} & \{$X_1$, $X_2$, $X_1$$X_2$\} & \{$Y_1$, $Y_2$, $Y_1$$Y_2$\} & \{$Z_1$$X_2$, $X_1$$Y_2$, $-Y_1$$Z_2$\} & \{$Z_1$$Y_2$, $X_1$$Z_2$, $-Y_1$$X_2$\}\\
2 & \{$Z_1$, $Z_2$, $Z_1$$Z_2$\} & \{$X_1$, $Y_2$, $X_1$$Y_2$\} & \{$Y_1$, $X_2$, $Y_1$$X_2$\} & \{$Z_1$$X_2$, $X_1$$Z_2$, $Y_1$$Y_2$\} & \{$Y_1$$Z_2$, $Z_1$$Y_2$, $X_1$$X_2$\}\\
3 & \{$X_1$, $X_2$, $X_1$$X_2$\} & \{$Y_1$, $Z_2$, $Y_1$$Z_2$\} & \{$Z_1$, $Y_2$, $Z_1$$Y_2$\} & \{$Z_1$$X_2$, $X_1$$Z_2$, $Y_1$$Y_2$\} & \{$X_1$$Y_2$, $Y_1$$X_2$, $Z_1$$Z_2$\}\\
4 & \{$Y_1$, $Y_2$, $Y_1$$Y_2$\} & \{$Z_1$, $X_2$, $Z_1$$X_2$\} & \{$X_1$, $Z_2$, $X_1$$Z_2$\} & \{$X_1$$Y_2$, $Y_1$$X_2$, $Z_1$$Z_2$\} & \{$Y_1$$Z_2$, $Z_1$$Y_2$, $X_1$$X_2$\}\\
5 & \{$Z_1$, $X_2$, $Z_1$$X_2$\} & \{$X_1$, $Y_2$, $X_1$$Y_2$\} & \{$Y_1$, $Z_2$, $Y_1$$Z_2$\} & \{$Z_1$$Z_2$, $X_1$$X_2$, $-Y_1$$Y_2$\} & \{$Z_1$$Y_2$, $X_1$$Z_2$, $-Y_1$$X_2$\}\\
6 & \{$Z_1$, $Y_2$, $Z_1$$Y_2$\} & \{$X_1$, $Z_2$, $X_1$$Z_2$\} & \{$Y_1$, $X_2$, $Y_1$$X_2$\} & \{$Z_1$$Z_2$, $X_1$$X_2$, $-Y_1$$Y_2$\} & \{$Z_1$$X_2$, $X_1$$Y_2$, $-Y_1$$Z_2$\}\\
\hline
\end{tabular}
\caption{The 6 sets of maximal MUBs made up of the 15 observables of a two-qubit system. Each row shows the five triads of observables making up a maximal MUB set, with any two rows have exactly one triad in common.}
\label{tab:MUBs} % is used to refer this table in the text
\end{table}

\begin{table}[ht]
\centering % used for centering table
\begin{tabular}{|c | c c c c |} % centered columns (4 columns)
\hline % inserts single horizontal line
Operators & ++  &   + --  &  -- +  &  -- --  \\
\hline
$Z_1$, $Z_2$, $Z_1$$Z_2$            & $1=1000$ & $2=0100$ & $3=0010$ & $4=0001$ \\
$X_1$, $X_2$, $X_1$$X_2$            & $5=1111$ & $6=1\overline{1}1\overline{1}$ & $7=11\overline{1}\overline{1}$ & $8=1\overline{1}\overline{1}1$ \\
$Y_1$, $Y_2$, $Y_1$$Y_2$            & $9=1ii\overline{1}$ & $10=1\overline{i}i1$ & $11=1i\overline{i}1$ & $12=1\overline{i}\overline{i}\overline{1}$ \\
$Z_1$, $X_2$, $Z_1$$X_2$            & $13=1100$ & $14=1\overline{1}00$ & $15=0011$ & $16=001\overline{1}$ \\
$X_1$, $Y_2$, $X_1$$Y_2$            & $17=1i1i$ & $18=1\overline{i}1\overline{i}$ & $19=1i\overline{1}\overline{i}$ & $20=1\overline{i}\overline{1}i$ \\
$Y_1$, $Z_2$, $Y_1$$Z_2$            & $21=10i0$ & $22=010i$ & $23=10\overline{i}0$ & $24=010\overline{i}$ \\
$Z_1$$X_2$, $X_1$$Z_2$, $Y_1$$Y_2$  & $25=111\overline{1}$ & $26=11\overline{1}1$ & $27=1\overline{1}11$ & $28=1\overline{1}\overline{1}\overline{1}$ \\
$X_1$$Y_2$, $Y_1$$X_2$, $Z_1$$Z_2$  & $29=100i$ & $30=01\overline{i}0$ & $31=01i0$ & $32=100\overline{i}$ \\
$Y_1$$Z_2$, $Z_1$$Y_2$, $X_1$$X_2$  & $33=1ii1$ & $34=1\overline{i}i\overline{1}$ & $35=1i\overline{i}\overline{1}$ & $36=1\overline{i}\overline{i}1$ \\
$Z_1$$Y_2$, $X_1$$Z_2$, $-Y_1$$X_2$ & $37=1i1\overline{i}$ & $38=1i\overline{1}i$ & $39=1\overline{i}1i$ & $40=1\overline{i}\overline{1}\overline{i}$ \\
\hline %inserts single line
\end{tabular}
\caption{The 40-ray system obtained by discarding the five triads of observables in the last row of Table \ref{tab:MUBs} and keeping only the eigenstates associated with the remaining triads. {\it Note that the numbering of the rays here is different from that in Table \ref{60States}}.}
\label{40States} % is used to refer this table in the text
\end{table}

The 40-40 system is small enough that all its parity proofs can be determined through an exhaustive computer search. A complete list of the  proofs in this system is shown in Table \ref{tab:40List}. An interesting feature of this list is that the proofs come in {\bf basis-complementary} pairs, i.e. pairs that have no bases in common and whose union gives back the entire 40-40 set. Tables \ref{tab:30-15}, \ref{tab:32-17} and \ref{tab:34-19} give one example of each of the pairs of basis-complementary proofs listed in Table \ref{tab:40List}. It is an interesting numerological observation that the total number of distinct parity proofs in this system, which is twice the total of the last column in Table \ref{tab:40List}, or 32768, can also be expressed as $2^{15}$, where 15 is the number of hybrid bases in each of the proofs. This resonates with a similar observation we made recently \cite{Waegell2011a}: the 24-24 Peres system has $512 = 2^9$ distinct parity proofs in it, where 9 is again the number of hybrid bases involved in the proofs.\\

\begin{table}[ht]
\centering % used for centering table
\begin{tabular}{|c |c |c |} % centered columns (3 columns)
\hline % inserts single horizontal line
Parity Proof & Complementary Proof & Number\\
\hline
30-15 ($30_{2}$-$15_{4}$) & 40-25 ($30_{2}10_{4}$-$25_{4}$) & 64  \\
32-17 ($30_{2}2_{4}$-$17_{4}$) &  38-23 ($30_{2}8_{4}$-$23_{4}$) &  2880 \\
34-19 ($30_{2}4_{4}$-$19_{4}$) &  36-21 ($30_{2}6_{4}$-$21_{4}$) & 13440   \\
\hline
\end{tabular}
\caption{Parity proofs in the 40-40 system. Each line shows a pair of basis-complementary proofs, by both their abbreviated and expanded symbols,   and indicates the number of replicas of each in the 40-40 set.}
\label{tab:40List} % is used to refer this table in the text
\end{table}

The five other 40-40 subsystems contain proofs that are identical to the ones above and related to them by symmetry. The role of symmetry in creating replicas of a basic proof pattern should now be clear. For example, the 30-15 proof shown in Table \ref{tab:30-15} has $32\times6 = 192$ replicas under symmetry in the full 60-105 set (on recalling that the 30-15 proofs come in two geometrically distinct types of equal numbers). In this case, and many of the others to be discussed, the amplification due to symmetry comes in two steps: first there is the replication due to the symmetries of the subsystem itself, and then there is a further amplification due to the symmetries that exchange that subsystem with other similar subsystems in the 60-105 set.\\

\begin{table}[ht]
\centering % used for centering table
\begin{tabular}{|| c c c c || c c c c | c c c c | c c c c ||} % centered columns (4 columns)
\hline\hline % inserts single horizontal line
1 & 2 & 3 & 4 & \textbf{1} & \textbf{2} & \textbf{15} & \textbf{16} & 6 & 8 & 13 & 15 & \textbf{17} & \textbf{20} & \textbf{31} & \textbf{32}\\
5 & 6 & 7 & 8 & \textbf{1} & \textbf{3} & \textbf{22} & \textbf{24} & 7 & 8 & 17 & 18 & 18 & 19 & 29 & 30\\
9 & 10 & 11 & 12 & 1 & 4 & 30 & 31 & \textbf{9} & \textbf{10} & \textbf{23} & \textbf{24} & 21 & 24 & 35 & 36\\
13 & 14 & 15 & 16 & \textbf{2} & \textbf{3} & \textbf{29} & \textbf{32} & 9 & 11 & 18 & 20 & \textbf{22} & \textbf{23} & \textbf{33} & \textbf{34}\\
17 & 18 & 19 & 20 & 2 & 4 & 21 & 23 & \textbf{9} & \textbf{12} & \textbf{26} & \textbf{27} & \textbf{25} & \textbf{27} & \textbf{38} & \textbf{40}\\
21 & 22 & 23 & 24 & 3 & 4 & 13 & 14 & 10 & 11 & 25 & 28 & 26 & 28 & 37 & 39\\
25 & 26 & 27 & 28 & \textbf{5} & \textbf{6} & \textbf{19} & \textbf{20} & \textbf{10} & \textbf{12} & \textbf{17} & \textbf{19} & \textbf{29} & \textbf{31} & \textbf{37} & \textbf{40}\\
29 & 30 & 31 & 32 & \textbf{5} & \textbf{7} & \textbf{14} & \textbf{16} & 11 & 12 & 21 & 22 & 30 & 32 & 38 & 39\\
33 & 34 & 35 & 36 & 5 & 8 & 34 & 35 & 13 & 16 & 27 & 28 & 33 & 35 & 39 & 40\\
37 & 38 & 39 & 40 & \textbf{6} & \textbf{7} & \textbf{33} & \textbf{36} & \textbf{14} & \textbf{15} & \textbf{25} & \textbf{26} & \textbf{34} & \textbf{36} & \textbf{37} & \textbf{38}\\
\hline\hline %inserts single line
\end{tabular}
\caption{A 30-15 proof (bold font) and its complementary 40-25 proof (ordinary font) in the 40-40 set of Fig. \ref{tab:40Bases}.}
\label{tab:30-15} % is used to refer this table in the text
\end{table}

\begin{table}[ht]
\centering % used for centering table
\begin{tabular}{|| c c c c || c c c c | c c c c | c c c c ||} % centered columns (4 columns)
\hline\hline % inserts single horizontal line
\textbf{1} & \textbf{2} & \textbf{3} & \textbf{4} & \textbf{1} & \textbf{2} & \textbf{15} & \textbf{16} & 6 & 8 & 13 & 15 & 17 & 20 & 31 & 32\\
\textbf{5} & \textbf{6} & \textbf{7} & \textbf{8} & \textbf{1} & \textbf{3} & \textbf{22} & \textbf{24} & 7 & 8 & 17 & 18 & \textbf{18} & \textbf{19} & \textbf{29} & \textbf{30}\\
9 & 10 & 11 & 12 & \textbf{1} & \textbf{4} & \textbf{30} & \textbf{31} & \textbf{9} & \textbf{10} & \textbf{23} & \textbf{24} & 21 & 24 & 35 & 36\\
13 & 14 & 15 & 16 & 2 & 3 & 29 & 32 & \textbf{9} & \textbf{11} & \textbf{18} & \textbf{20} & \textbf{22} & \textbf{23} & \textbf{33} & \textbf{34}\\
17 & 18 & 19 & 20 & 2 & 4 & 21 & 23 & 9 & 12 & 26 & 27 & 25 & 27 & 38 & 40\\
21 & 22 & 23 & 24 & 3 & 4 & 13 & 14 & \textbf{10} & \textbf{11} & \textbf{25} & \textbf{28} & \textbf{26} & \textbf{28} & \textbf{37} & \textbf{39}\\
25 & 26 & 27 & 28 & \textbf{5} & \textbf{6} & \textbf{19} & \textbf{20} & 10 & 12 & 17 & 19 & \textbf{29} & \textbf{31} & \textbf{37} & \textbf{40}\\
29 & 30 & 31 & 32 & \textbf{5} & \textbf{7} & \textbf{14} & \textbf{16} & 11 & 12 & 21 & 22 & 30 & 32 & 38 & 39\\
33 & 34 & 35 & 36 & \textbf{5} & \textbf{8} & \textbf{34} & \textbf{35} & 13 & 16 & 27 & 28 & \textbf{33} & \textbf{35} & \textbf{39} & \textbf{40}\\
37 & 38 & 39 & 40 & 6 & 7 & 33 & 36 & \textbf{14} & \textbf{15} & \textbf{25} & \textbf{26} & 34 & 36 & 37 & 38\\
\hline\hline %inserts single line
\end{tabular}
\caption{A 32-17 proof (bold font) and its complementary 38-23 proof (ordinary font) in the 40-40 set of Fig. \ref{tab:40Bases}.}
\label{tab:32-17} % is used to refer this table in the text
\end{table}

\begin{table}[ht]
\centering % used for centering table
\begin{tabular}{|| c c c c || c c c c | c c c c | c c c c ||} % centered columns (4 columns)
\hline\hline % inserts single horizontal line
\textbf{1} & \textbf{2} & \textbf{3} & \textbf{4} & \textbf{1} & \textbf{2} & \textbf{15} & \textbf{16} & 6 & 8 & 13 & 15 & 17 & 20 & 31 & 32\\
\textbf{5} & \textbf{6} & \textbf{7} & \textbf{8} & \textbf{1} & \textbf{3} & \textbf{22} & \textbf{24} & 7 & 8 & 17 & 18 & \textbf{18} & \textbf{19} & \textbf{29} & \textbf{30}\\
\textbf{9} & \textbf{10} & \textbf{11} & \textbf{12} & \textbf{1} & \textbf{4} & \textbf{30} & \textbf{31} & \textbf{9} & \textbf{10} & \textbf{23} & \textbf{24} & 21 & 24 & 35 & 36\\
13 & 14 & 15 & 16 & 2 & 3 & 29 & 32 & \textbf{9} & \textbf{11} & \textbf{18} & \textbf{20} & \textbf{22} & \textbf{23} & \textbf{33} & \textbf{34}\\
17 & 18 & 19 & 20 & 2 & 4 & 21 & 23 & \textbf{9} & \textbf{12} & \textbf{26} & \textbf{27} & \textbf{25} & \textbf{27} & \textbf{38} & \textbf{40}\\
21 & 22 & 23 & 24 & 3 & 4 & 13 & 14 & 10 & 11 & 25 & 28 & 26 & 28 & 37 & 39\\
25 & 26 & 27 & 28 & \textbf{5} & \textbf{6} & \textbf{19} & \textbf{20} & 10 & 12 & 17 & 19 & \textbf{29} & \textbf{31} & \textbf{37} & \textbf{40}\\
29 & 30 & 31 & 32 & \textbf{5} & \textbf{7} & \textbf{14} & \textbf{16} & 11 & 12 & 21 & 22 & 30 & 32 & 38 & 39\\
33 & 34 & 35 & 36 & \textbf{5} & \textbf{8} & \textbf{34} & \textbf{35} & 13 & 16 & 27 & 28 & \textbf{33} & \textbf{35} & \textbf{39} & \textbf{40}\\
\textbf{37} & \textbf{38} & \textbf{39} & \textbf{40} & 6 & 7 & 33 & 36 & \textbf{14} & \textbf{15} & \textbf{25} & \textbf{26} & 34 & 36 & 37 & 38\\
\hline\hline %inserts single line
\end{tabular}
\caption{A 34-19 proof (bold font) and its complementary 36-21 proof (ordinary font) in the 40-40 set of Fig. \ref{tab:40Bases}.}
\label{tab:34-19} % is used to refer this table in the text
\end{table}

\begin{figure}[htp]
\begin{center}
\includegraphics[width=.60\textwidth]{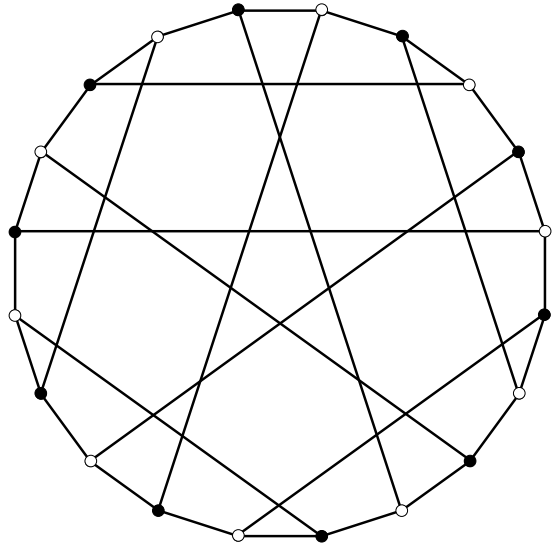}
\end{center}
\caption{Levi graph of the Desarguesian configuration, $10_{3}$. The solid and open circles represent two sets of ten objects. Each object of either set is associated with the three objects of the other to which it is connected by line segments.}
\label{Desarg}
\end{figure}

Though the proofs in Table \ref{tab:40List} were discovered through a computer search, we later found simple constructions for them based on deletions of selected subsets of rays from the 40-40 system. The most interesting of these constructions is the one for the 30-15 proof, which we now describe. Let us define a {\bf Desarguesian configuration} as any two sets of ten objects with the property that each object of either set is associated with three objects of the other. The original example of such a configuration was provided by Desargues \cite{Hilbert}, who showed that if two triangles are perspective from a point (i.e. the joins of corresponding vertices pass through a point), then they are also perspective from a line (i.e. the extensions of their corresponding sides intersect in three points that lie on a line). The two sets of objects in this case are ten points (namely, the vertices of the two triangles, the perspective point and the three points in which pairs of their sides intersect) and ten lines (namely, the sides of the two triangles, the lines joining corresponding vertices to the perspective point and the perspective line), and the association between the points and the lines is one of incidence. Because the ten objects of either set are each associated with three of the other, the symbol $10_{3}$ is sometimes used to denote a Desarguesian configuration. The English mathematician Cayley gave another example of a Desarguesian configuration: he pointed out \cite{Coxeter} that the ten lines and ten planes determined by five points of general position in projective 3-space meet an arbitrary plane in the ten points and ten lines of a Desarguesian configuration. A geometrical model of a Desarguesian configuration is provided by a 20-gon whose alternate vertices represent the two sets of ten objects, with the edges and certain diagonals representing the associations between the objects of the two sets (see Fig. \ref{Desarg}).\\

The 40-40 system contains two types of Desarguesian configurations that hold the key to the construction of the 30-15 proofs. Let us call the rays of the 40-40 system ``points" and define a ``line" as any set of three "points" (i.e. rays) with the property that the components of one can be expressed as a linear combination of those of the other two. With these definitions it is straightforward to check that the 40-40 system contains 32 Desarguesian configurations of ``points" and ``lines", with each ``point" being incident with three ``lines" of its configuration and vice-versa. One example of such a configuration is provided by the ten rays (or ``points") 4,8,11,13,18,21,28,30,35 and 39 and the ten lines (4,8,28), (4,11,35), (4,18,39), (8,11,30), (8,21,39), (11,13,39), (13,18,35), (13,21,30), (18,21,28) and (28,30,35), where each line has been indicated by the three "points" on it. If all the bases containing any of the foregoing rays are omitted from the 40 bases of Fig. \ref{tab:40Bases}, the remaining 15 bases give the 30-15 proof of Table \ref{tab:30-15}. A second type of Desarguesian configuration is obtained by defining the ``points" as before but replacing the ``lines" with ``triangles", where a ``triangle" is defined as any set of three rays that are mutually unbiased (i.e., the squared modulus of the overlap of any two normalized rays is 1/4) but nevertheless do not lie on a ``line". An example of a ``triangle" is given by the rays 1,5 and 27. There are exactly 32 Desarguesian configurations of this second kind in a 40-40 set, and omitting all bases involving any of the rays in one of them from the full 40-40 set again leads to a 30-15 proof (of a geometrically different kind from the earlier one). Thus we have shown how all the 30-15 proofs in a 40-40 system can be obtained by dropping bases picked out by one of the two types of Desarguesian configurations in it.

 \section{\label{sec:36-36}The 36-36 subsystem and its parity proofs}

\begin{figure}[htp]
\begin{center}
\includegraphics[width=.60\textwidth]{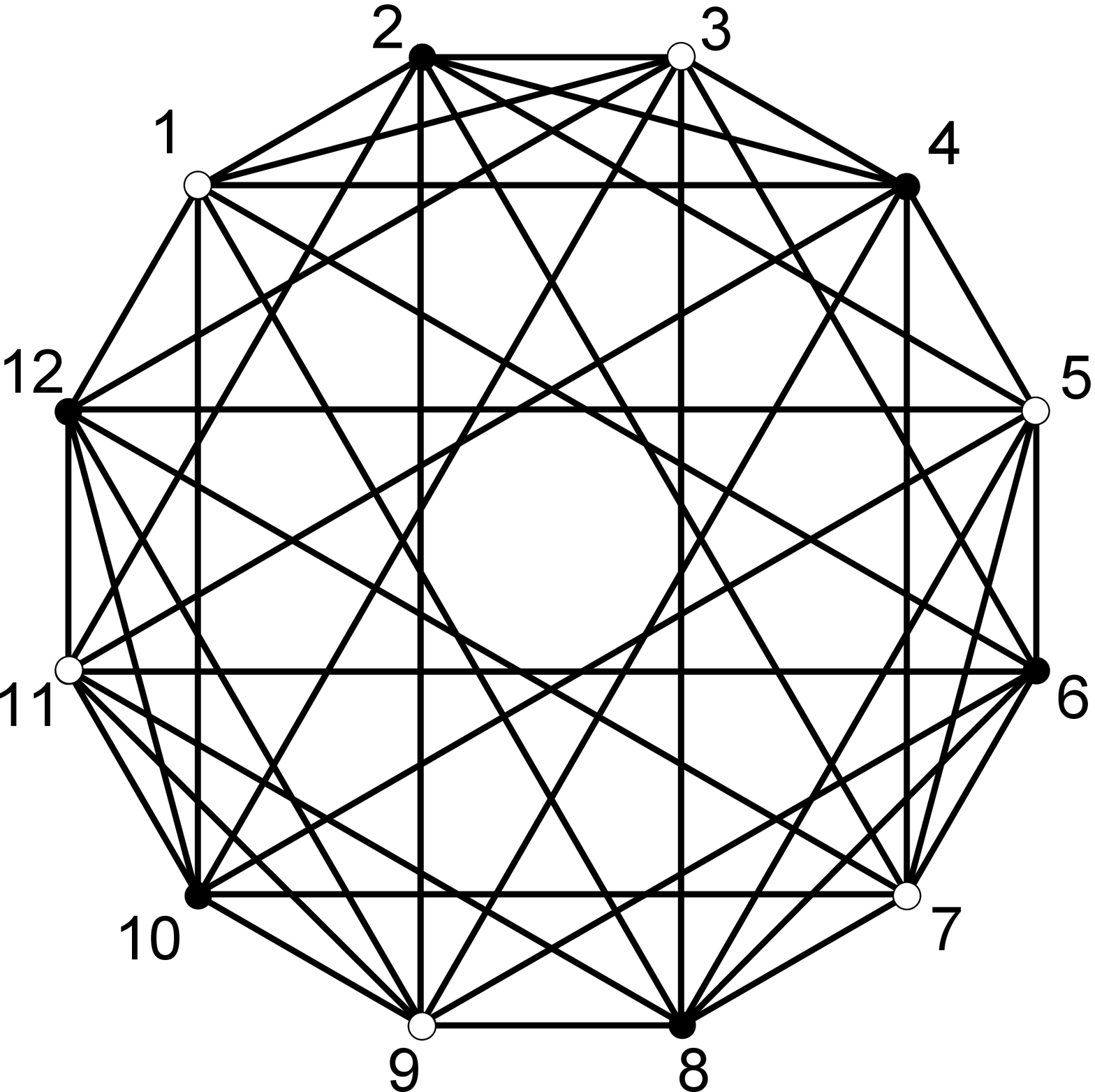}
\end{center}
\caption{Kochen-Specker diagram of the 12 rays in a ``dodecagon". The rays are shown as filled and open circles at the vertices of a dodecagon and numbered from 1 to 12. Each ray is orthogonal to seven others, six of the opposite type and one of the same type. The rays form the three (mutually unbiased) pure bases 1 2 3 4, 5 6 7 8 and 9 10 11 12 and the six hybrid bases 1 3 6 8, 1 3 10 12, 5 7 2 4, 5 7 10 12, 9 11 2 4 and 9 11 4 6. The pure bases consist of consecutive sets of four vertices along the perimeter of the 12-gon, while the hybrid bases are made up of a pair of open circles from one pure basis and a pair of filled circles from another. This configuration can be characterized by the symbol $12_{3}$-$9_{4}$, because each of the twelve rays occurs in three of the nine bases. If the rays in each of the rows of Table \ref{12-gonTablesA} are arranged at the vertices of this dodecagon, beginning at the vertex marked 1 and proceeding clockwise, their orthogonalities are represented faithfully by the lines in the figure and the nine bases formed by them can be picked out by replacing the numbers 1-12 in the above listings by the rays at those positions.}
\label{KS12-gon}
\end{figure}

 The 60 rays possess several subsets of 12 rays with the property that each ray is orthogonal to seven others in the group. Table \ref{12-gonTablesA} shows the 15 such sets of rays that exist within the 60 rays. We will term any such set of 12 rays a {\bf dodecagon} because its Kochen-Specker diagram, shown in Fig. \ref{KS12-gon}, has the form of a dodecagon with all its edges and thirty of its diagonals drawn in. Table \ref{12-gonTablesB} shows the six different coverings of all 60 rays by sets of five non-overlapping dodecagons. If one keeps all the rays in any three dodecagons in any row of Table \ref{12-gonTablesB}, one gets 36 rays, and if one picks out all the hybrid bases formed by these rays one gets a 36-36 system. The total number of 36-36 systems that can be constructed in this way is $10\times6 = 60$.\\

\begin{table}[ht]
\centering % used for centering table
\begin{tabular}{||c | c c c c c c c c c c c c |} % centered columns (4 columns)
\hline % inserts single horizontal line
Index & \multicolumn{12}{|c|}{12 rays in a dodecagon}  \\
\hline
1 & 1 & 3 & 2 & 4 & 13 & 15 & 14 & 16 & 25 & 27 & 26 & 28\\
2 & 1 & 2 & 3 & 4 & 21 & 22 & 23 & 24 & 29 & 30 & 31 & 32\\
3 & 1 & 2 & 4 & 3 & 41 & 42 & 44 & 43 & 49 & 51 & 50 & 52\\
4 & 5 & 7 & 6 & 8 & 17 & 19 & 18 & 20 & 29 & 31 & 30 & 32\\
5 & 5 & 6 & 7 & 8 & 13 & 14 & 15 & 16 & 33 & 34 & 35 & 36\\
6 & 5 & 6 & 8 & 7 & 45 & 46 & 48 & 47 & 49 & 50 & 51 & 52\\
7 & 9 & 11 & 10 & 12 & 21 & 23 & 22 & 24 & 33 & 35 & 34 & 36\\
8 & 9 & 10 & 11 & 12 & 17 & 18 & 19 & 20 & 25 & 26 & 27 & 28\\
9 & 9 & 10 & 12 & 11 & 37 & 38 & 40 & 39 & 50 & 49 & 51 & 52\\
10 & 13 & 14 & 16 & 15 & 37 & 39 & 38 & 40 & 53 & 55 & 54 & 56\\
11 & 17 & 18 & 20 & 19 & 41 & 43 & 42 & 44 & 53 & 54 & 55 & 56\\
12 & 21 & 22 & 24 & 23 & 45 & 47 & 46 & 48 & 54 & 53 & 55 & 56\\
13 & 25 & 26 & 28 & 27 & 45 & 46 & 47 & 48 & 57 & 59 & 58 & 60\\
14 & 29 & 30 & 32 & 31 & 37 & 38 & 39 & 40 & 57 & 58 & 59 & 60\\
15 & 33 & 34 & 36 & 35 & 41 & 42 & 43 & 44 & 58 & 57 & 59 & 60\\
\hline\hline %inserts single line
\end{tabular}
\caption{The 15 ``dodecagons" in the set of 60 rays.}
\label{12-gonTablesA} % is used to refer this table in the text
\end{table}

\begin{table}[ht]
\centering % used for centering table
\begin{tabular}{||c | c c c c c |} % centered columns (4 columns)
\hline % inserts single horizontal line
Index & \multicolumn{5}{|c|}{Covering}  \\
\hline
1 & 1 & 4 & 9 & 12 & 15\\
2 & 1 & 6 & 7 & 11 & 14\\
3 & 2 & 5 & 9 & 11 & 13\\
4 & 2 & 6 & 8 & 10 & 15\\
5 & 3 & 4 & 7 & 10 & 13\\
6 & 3 & 5 & 8 & 12 & 14\\
\hline\hline %inserts single line
\end{tabular}
\caption{The six coverings of the 60 rays by five non-overlapping dodecagons, with the dodecagons numbered as in Table \ref{12-gonTablesA}.}
\label{12-gonTablesB} % is used to refer this table in the text
\end{table}

The 36-36 system (like the 40-40 system) is small enough to allow all its parity proofs to be determined through an exhaustive computer search. The complete list of proofs in this system is 18-9, 22-11, 24-13, 26-13, 26-15, 28-15 (34-21), 29-15 (35-21), 30-15, 30-17 (32-19), 31-17 (33-19), 32-17 (34-19), 32-19 (30-17), 33-19 (31-17), 34-19 (32-17), 34-21 (28-15), 35-21 (29-15) and 36-21, where we have used the abbreviated symbols for the proofs and indicated the basis-complementary proof to a given one after it in brackets, if it is basis-critical. Since the 36-36 system has an even number of bases, the basis-complement of any parity proof contained in it is automatically another parity proof. However the complement need not be basis-critical, and then we would discard it. This explains why every proof in the list is not accompanied by its basis-complement. We should also point out that the first proof in this list, 18-9, is actually one of the six different types of parity proofs in the 24-24 Peres set. However all the other proofs are new.\\

Tables \ref{36States} and \ref{BasisTable36-36} show an example of a 36-36 set and a pair of basis-complementary parity proofs in it.

\begin{table}[ht]
\centering % used for centering table
\begin{tabular}{|c | c c c c |} % centered columns (4 columns)
\hline % inserts single horizontal line
Operators & ++  &   + --  &  -- +  &  -- --  \\
\hline
$Z_1$, $Z_2$, $Z_1$$Z_2$            & $1=1000$ & $2=0100$ & $3=0010$ & $4=0001$ \\
$X_1$, $X_2$, $X_1$$X_2$            & $5=1111$ & $6=1\overline{1}1\overline{1}$ & $7=11\overline{1}\overline{1}$ & $8=1\overline{1}\overline{1}1$ \\
$Y_1$, $Y_2$, $Y_1$$Y_2$            & $9=1ii\overline{1}$ & $10=1\overline{i}i1$ & $11=1i\overline{i}1$ & $12=1\overline{i}\overline{i}\overline{1}$ \\
$Z_1$, $X_2$, $Z_1$$X_2$            & $13=1100$ & $14=1\overline{1}00$ & $15=0011$ & $16=001\overline{1}$ \\
$X_1$, $Y_2$, $X_1$$Y_2$            & $17=1i1i$ & $18=1\overline{i}1\overline{i}$ & $19=1i\overline{1}\overline{i}$ & $20=1\overline{i}\overline{1}i$ \\
$Z_1$, $Y_2$, $Z_1$$Y_2$            & $21=1i00$ & $22=1\overline{i}00$ & $23=001i$ & $24=001\overline{i}$ \\
$X_1$, $Z_2$, $X_1$$Z_2$            & $25=1010$ & $26=0101$ & $27=10\overline{1}0$ & $28=010\overline{1}$ \\
$Z_1$$X_2$, $X_1$$Z_2$, $Y_1$$Y_2$  & $29=111\overline{1}$ & $30=11\overline{1}1$ & $31=1\overline{1}11$ & $32=1\overline{1}\overline{1}\overline{1}$ \\
$Z_1$$Z_2$, $X_1$$X_2$, $-Y_1$$Y_2$ & $33=1001$ & $34=100\overline{1}$ & $35=0110$ & $36=01\overline{1}0$ \\
\hline
\end{tabular}
\caption{The 36 rays obtained by keeping all rays in the first three dodecagons in the first row of Table \ref{12-gonTablesB}. The rays form nine pure bases, corresponding to the triads of observables shown in the first column. {\it Note that the numbering of the rays here is different from that in the earlier sections.}}
\label{36States} % is used to refer this table in the text
\end{table}

\begin{table}[ht]
\centering % used for centering table
\begin{tabular}{|c c c c | c c c c | c c c c | c c c c |} % centered columns (4 columns)
\hline % inserts single horizontal line
\textbf{1} & \textbf{2} & \textbf{15} & \textbf{16} & \textbf{5} & \textbf{6} & \textbf{27} & \textbf{28} & 9 & 12 & 30 & 31 & 15 & 16 & 21 & 22\\
\textbf{1} & \textbf{2} & \textbf{23} & \textbf{24} & \textbf{5} & \textbf{7} & \textbf{14} & \textbf{16} & 9 & 12 & 33 & 36 & 17 & 18 & 27 & 28\\
\textbf{1} & \textbf{3} & \textbf{26} & \textbf{28} & \textbf{5} & \textbf{8} & \textbf{34} & \textbf{36} & 10 & 11 & 29 & 32 & \textbf{17} & \textbf{19} & \textbf{22} & \textbf{24}\\
\textbf{1} & \textbf{4} & \textbf{35} & \textbf{36} & 6 & 7 & 33 & 35 & 10 & 11 & 34 & 35 & \textbf{18} & \textbf{20} & \textbf{21} & \textbf{23}\\
2 & 3 & 33 & 34 & 6 & 8 & 13 & 15 & 10 & 12 & 17 & 19 & 19 & 20 & 25 & 26\\
2 & 4 & 25 & 27 & \textbf{7} & \textbf{8} & \textbf{17} & \textbf{18} & 10 & 12 & 21 & 23 & 25 & 28 & 30 & 32\\
3 & 4 & 13 & 14 & 7 & 8 & 25 & 26 & 13 & 14 & 23 & 24 & \textbf{26} & \textbf{27} & \textbf{29} & \textbf{31}\\
\textbf{3} & \textbf{4} & \textbf{21} & \textbf{22} & 9 & 11 & 18 & 20 & 13 & 16 & 31 & 32 & 29 & 32 & 33 & 36\\
\textbf{5} & \textbf{6} & \textbf{19} & \textbf{20} & 9 & 11 & 22 & 24 & \textbf{14} & \textbf{15} & \textbf{29} & \textbf{30} & \textbf{30} & \textbf{31} & \textbf{34} & \textbf{35}\\
\hline
\end{tabular}
\caption{The 36 hybrid bases formed by the 36 rays of Table \ref{36States}. Each ray occurs in four bases, so this is a $36_{4}$-$36_{4}$ system. A  $26_{2}2_{4}$-$15_{4}$ parity proof is shown in bold, while the remaining bases make up the basis-complementary $26_{2}8_{4}$-$21_{4}$ proof.}
\label{BasisTable36-36} % is used to refer this table in the text
\end{table}

 \section{\label{sec:Full} Other proofs in the 60-105 system}

So far we have identified two subsystems of the 60-105 system, namely the 40-40 and 36-36 systems, and given a complete account of the parity proofs in them. The only smaller subsystem of interest is the 24-24 Peres system, whose parity proofs have been completely mapped out \cite{Waegell2011a}. Some larger subsystems of the 60-105 system that are of interest are the 36-45 system (obtained from a 36-36 system by adding on the 9 pure bases formed by the 36 rays) and the 48-60 and 48-72 systems (obtained by keeping the rays in any four dodecagons in any row of Table \ref{12-gonTablesB} together with just the hybrid bases formed by them, or both the hybrid and pure bases formed by them). We have explored these larger subsystems and found a huge number of new proofs in them. The search for these new proofs proved more time consuming because a larger number of bases had to be explored and rays with multiplicities greater than two also occurred more frequently among the proofs. The full 60-105 system is of course the most difficult one to mine, for the various reasons mentioned, and also because of the very large number of non-critical proofs that are netted and have to be weeded out. The strategy of starting with the smallest subsystems and then working upwards, which we have adopted in this work, thus seems the most fruitful approach.\\

We give below a small sample of some of the other proofs we have found in the 60-105 system, grouped according to the number of bases in the proof. \\

{\indent} 19 bases: {\space} $35_{2}1_{6}$,{\space \space}$33_{2}1_{4}1_{6}$,{\space \space}$31_{2}2_{4}1_{6}$,{\space \space}$29_{2}3_{4}1_{6}$\\ \\
{\indent} 21 bases: {\space} $39_{2}1_{6}$,{\space \space}$37_{2}1_{4}1_{6}$,{\space \space}$35_{2}2_{4}1_{6}$,{\space \space}$33_{2}3_{4}1_{6}$,{\space \space}$31_{2}4_{4}1_{6}$,{\space \space}$29_{2}5_{4}1_{6}${\space \space}\\ \\
{\indent} 23 bases: {\space} $43_{2}1_{6}$,{\space \space}$41_{2}1_{4}1_{6}$,{\space \space}$39_{2}2_{4}1_{6}$,{\space \space}$37_{2}3_{4}1_{6}$,{\space \space}$35_{2}4_{4}1_{6}$,{\space \space}$29_{2}7_{4}1_{6}$\\ \\
{\indent} 29 bases: {\space} $36_{2}11_{4}$,{\space \space}$37_{2}9_{4}1_{6}$,{\space \space}$38_{2}7_{4}2_{6}$,{\space \space}$39_{2}5_{4}3_{6}$,{\space \space}$40_{2}3_{4}4_{6}$\\ \\
{\indent} 31 bases: {\space} $42_{2}10_{4}$,{\space \space}$43_{2}8_{4}1_{6}$,{\space \space}$44_{2}6_{4}2_{6}$,{\space \space}$45_{2}4_{4}3_{6}$,{\space \space}$46_{2}2_{4}4_{6}$\\ \\

As one goes from left to right in any row of the above listing, rays of multiplicity 2 get traded for a smaller number of rays of multiplicities 4 and 6. The appearance of rays of multiplicity 6 in many of these proofs is a new feature that was not present in the 40-40 and 36-36 subsystems. Table \ref{tab:47-29} shows an example of the last proof in the fourth line above. The proofs in the fourth and fifth lines illustrate an interesting phenomenon we term {\bf isomerism}. In chemistry, isomers are compounds that have the same chemical formula but different structural formulas. Analogously, we term two parity proofs isomers if they have the same abbreviated symbol but different expanded symbols, reflecting the fact that their structures are different. For example, all the proofs in the fourth line have the abbreviated symbol 47-29, but they involve different numbers of rays of multiplicities 2, 4 and 6. Similarly, the proofs in the fifth line are all 52-31 proofs but again differ in their multiplicities. Another way of stating the difference between isomers is that they are unitarily inequivalent to each other and thus geometrically distinct (unlike the various replicas of a given proof under symmetry). Isomerism can be more subtle than in the examples just discussed. The 60-105 system has a 30-15 parity proof involving 30 rays that each occur twice. The 600-cell \cite{Waegell2010} also has a 30-15 proof involving 30 rays that occur twice each. However these two proofs are not identical but are isomers of each other. The simplest way of seeing this is to note that all the rays in the latter proof are real, whereas there is no choice of basis that will allow all the rays in the former proof to be made real. We have found hundreds of examples of isomers among all the proofs we have found. Isomerism is experimentally significant, because different experimental arrangements would be needed to test proofs that are isomers of each other. \\

\begin{table}[ht]
\centering % used for centering table
\begin{tabular}{|c c c c | c c c c | c c c c | c c c c | c c c c|} % centered columns (4 columns)
\hline % inserts single horizontal line
1 & 2 & 3 & 4 & 5 & 6 & 7 & 8 & 9 & 10 & 11 & 12 & 13 & 14 & 15 & 16 & 17 & 18 & 19 & 20 \\
21 & 22 & 23 & 24 & 25 & 26 & 27 & 28 & 29 & 30 & 31 & 32 & 33 & 34 & 35 & 36 & 37 & 38 & 39 & 40 \\
41 & 42 & 43 & 44 & 45 & 46 & 47 & 48 & 49 & 50 & 51 & 52 & 53 & 54 & 55 & 56 & 57 & 58 & 59 & 60 \\
\hline
\textbf{1} & \textbf{2} & \textbf{15} & \textbf{16} & \textbf{1} & \textbf{2} & \textbf{27} & \textbf{28} & \textbf{1} & \textbf{3} & \textbf{22} & \textbf{24} & 1 & 3 & 30 & 32 & \textbf{1} & \textbf{4} & \textbf{42} & \textbf{43} \\
1 & 4 & 51 & 52 & 2 & 3 & 41 & 44 & 2 & 3 & 49 & 50 & 2 & 4 & 21 & 23 & 2 & 4 & 29 & 31 \\
\textbf{3} & \textbf{4} & \textbf{13} & \textbf{14} & 3 & 4 & 25 & 26 & \textbf{5} & \textbf{6} & \textbf{19} & \textbf{20} & \textbf{5} & \textbf{6} & \textbf{31} & \textbf{32} & \textbf{5} & \textbf{7} & \textbf{14} & \textbf{16} \\
\textbf{5} & \textbf{7} & \textbf{34} & \textbf{36} & \textbf{5} & \textbf{8} & \textbf{46} & \textbf{47} & \textbf{5} & \textbf{8} & \textbf{50} & \textbf{52} & \textbf{6} & \textbf{7} & \textbf{45} & \textbf{48} & \textbf{6} & \textbf{7} & \textbf{49} & \textbf{51} \\
\textbf{6} & \textbf{8} & \textbf{13} & \textbf{15} & \textbf{6} & \textbf{8} & \textbf{33} & \textbf{35} & \textbf{7} & \textbf{8} & \textbf{17} & \textbf{18} & \textbf{7} & \textbf{8} & \textbf{29} & \textbf{30} & 9 & 10 & 23 & 24 \\
\textbf{9} & \textbf{10} & \textbf{35} & \textbf{36} & 9 & 11 & 18 & 20 & 9 & 11 & 26 & 28 & 9 & 12 & 38 & 39 & \textbf{9} & \textbf{12} & \textbf{49} & \textbf{52} \\
10 & 11 & 37 & 40 & 10 & 11 & 50 & 51 & \textbf{10} & \textbf{12} & \textbf{17} & \textbf{19} & 10 & 12 & 25 & 27 & 11 & 12 & 21 & 22 \\
11 & 12 & 33 & 34 & 13 & 14 & 27 & 28 & 13 & 15 & 34 & 36 & 13 & 16 & 39 & 40 & 13 & 16 & 55 & 56 \\
14 & 15 & 37 & 38 & 14 & 15 & 53 & 54 & 14 & 16 & 33 & 35 & 15 & 16 & 25 & 26 & 17 & 18 & 31 & 32 \\
17 & 19 & 26 & 28 & 17 & 20 & 43 & 44 & 17 & 20 & 54 & 56 & 18 & 19 & 41 & 42 & 18 & 19 & 53 & 55 \\
\textbf{18} & \textbf{20} & \textbf{25} & \textbf{27} & 19 & 20 & 29 & 30 & 21 & 22 & 35 & 36 & 21 & 23 & 30 & 32 & 21 & 24 & 47 & 48 \\
21 & 24 & 53 & 56 & \textbf{22} & \textbf{23} & \textbf{45} & \textbf{46} & 22 & 23 & 54 & 55 & 22 & 24 & 29 & 31 & \textbf{23} & \textbf{24} & \textbf{33} & \textbf{34} \\
25 & 28 & 46 & 48 & \textbf{25} & \textbf{28} & \textbf{59} & \textbf{60} & 26 & 27 & 45 & 47 & 26 & 27 & 57 & 58 & 29 & 32 & 38 & 40 \\
\textbf{29} & \textbf{32} & \textbf{58} & \textbf{60} & 30 & 31 & 37 & 39 & \textbf{30} & \textbf{31} & \textbf{57} & \textbf{59} & 33 & 36 & 42 & 44 & 33 & 36 & 57 & 60 \\
34 & 35 & 41 & 43 & 34 & 35 & 58 & 59 & 37 & 38 & 55 & 56 & 37 & 39 & 58 & 60 & 37 & 40 & 49 & 52 \\
38 & 39 & 50 & 51 & 38 & 40 & 57 & 59 & 39 & 40 & 53 & 54 & 41 & 42 & 54 & 56 & 41 & 43 & 57 & 60 \\
41 & 44 & 51 & 52 & \textbf{42} & \textbf{43} & \textbf{49} & \textbf{50} & 42 & 44 & 58 & 59 & 43 & 44 & 53 & 55 & 45 & 46 & 53 & 56 \\
45 & 47 & 59 & 60 & 45 & 48 & 50 & 52 & \textbf{46} & \textbf{47} & \textbf{49} & \textbf{51} & \textbf{46} & \textbf{48} & \textbf{57} & \textbf{58} & 47 & 48 & 54 & 55 \\
\hline
\end{tabular}
\caption{A $40_{2}3_{4}4_{6}$-$29_{4}$ parity proof is shown in bold within the bases of the 60-105 system. Rays 1,46,49 have multiplicity four, rays 5,6,7,8 have multiplicity six, and all the remaining rays have multiplicity two.}
\label{tab:47-29} % is used to refer this table in the text
\end{table}

 \section{\label{sec:Discussion} Discussion}

 We have shown that the 60-105 system of rays and bases connected with a pair of qubits has a cornucopia of parity proofs of the KS theorem in it. The proofs involve anywhere from 9 to 41 bases (with all odd numbers in this range being possible), with several different numbers of rays generally being possible for each basis size. Each proof can be characterized most precisely through its expanded symbol, which specifies not just the number of rays and bases involved in the proof but the multiplicities of the rays as well. However we have found that even the expanded symbol sometimes fails to characterize a proof completely because different proofs with the same symbol could be unitarily inequivalent. The number of unitarily inequivalent parity proofs runs into the tens of thousands, and we have not attempted a systematic count. Each distinct type of proof typically has many replicas (sometimes thousands) under the symmetries of the system. All these effects conspire to make the total number of distinct parity proofs contained in the 60-105 system an astronomically large number. We find this number hard to estimate but would guess that it is somewhere in the vicinity of a thousand million. We have exhibited only a handful of proofs in this paper, but hope to make a much wider sample available on a website we plan to set up.\\

 It is a matter of some astonishment to us that a set of 105 bases, which can be fitted easily into a page, should contain something like $10^{9}$ parity proofs in it. This ratio of parity proofs to bases is the largest in any system we are aware of. We recently investigated a 60-75 system derived from the 600-cell \cite{Waegell2010} and found about 100 million parity proofs in it. That seems like the closest competitor to the present system, but we believe the present one wins out. What makes both these systems attractive is the combination of the large number of parity proofs in them together with the fact that every one of the proofs is so transparent. It is worth pointing out that both these 60-ray systems contain many basis-critical KS sets that are not parity proofs. These proofs are not as transparent as parity proofs, but they are just as conclusive. A distinguishing feature of these proofs is that they can involve an even number of bases, whereas parity proofs necessarily involve an odd number. In a recent paper \cite{Megill2011} we estimated the total number of critical KS sets of all types (both parity proofs and others) in the 60-75 system to be on the order of $10^{12}$. We think the total number of sets in the present system is just as large, but that still remains to be confirmed.\\

 We pointed out at the beginning that the 15 observables connected with a pair of qubits also form 15 triads of commuting observables. This symmetry is further enhanced by the fact that each observable occurs in three triads and each triad involves three observables. In other words, the observables and the triads form a $15_{3}$ configuration, i.e., two sets of 15 objects with the property that each object of either set is associated with three objects of the other. Geometers have identified a variety of examples of $15_{3}$ configurations. One of the simplest is known as a {\it hexastigm} because it is based on the numbers 1 to 6. Let a {\it duad} be an unordered pair of numbers from this set and a {\it syntheme} a set of three duads that include all six numbers between them. There are 15 duads and 15 synthemes and each is incident with three members of the opposite kind. Coxeter \cite{Coxeter} arranges the synthemes in a 6 x 6 table symmetrical about its diagonal (and with no entries along the diagonal) in such a way that every row and every column contains five synthemes that include all 15 duads among them. This table can be made to pass into a symmetrical version of our Table \ref{tab:MUBs} (showing the maximal sets of MUBs in a two-qubit system) if one translates the duads and synthemes into observables and triads of observables in a fairly obvious fashion. We have not been able to push this analogy to obtain further insights into the two-qubit system or the KS proofs contained in it, but we nevertheless find it striking enough to be worth mentioning.\\

An alternative approach \cite{Saniga} to the two-qubit Pauli group exploits its connection with the geometry of the generalized quadrangle GQ(2,2). This approach allows many significant properties of the observables, such as the sets of MUBs shown in Table \ref{tab:MUBs}, to be derived from purely geometrical considerations. These ideas can also be generalized, to some extent, to a system of several qudits \cite{Planat}.\\

There is one application of Table \ref{tab:MUBs} to quantum tomography, mentioned by one of us in an earlier work \cite{Aravind2006}, that is worth recalling. Following the seminal work of Wootters and others \cite{Wootters}, it is known that the most efficient method of determining the 15 parameters of an arbitrary two-qubit state is by carrying out measurements of five triads of observables that constitute a maximal set of MUBs. The triads that one uses for this purpose can be the ones in any of the rows of Table \ref{tab:MUBs}. However a superior strategy would be to measure all 15 triads instead of just the five in a particular row. This would involve three times as much work, but would allow the state reconstruction to be done in six different ways based on the rows of Table \ref{tab:MUBs}, and so lead to a two-to-one return on the investment. \\

{\bf Acknowledgements.} A part of this work was presented by one of us (PKA) at the Workshop on Topological Quantum Information held in Pisa, Italy, during May 16-17, 2011. PKA would like to thank the organizers, Professors Lou Kauffman and Sam Lomonaco, for the invitation to participate in the workshop and for the fruitful discussions that took place with them and the other participants during the visit. Thanks are also due to the Centro di Ricerca Matematica for its hospitality and support during the workshop. \\

%\bibliographystyle{spphys}
%\bibliography{Peres24}

\end{document}